\documentclass[aps, prd, preprint, amsfonts,
  amssymb, amsmath, showpacs, a4paper,nofootinbib,superscriptaddress]{revtex4-1}  
\usepackage[T1]{fontenc} 

\usepackage{slashed}
\usepackage{braket}

\newcommand{\be}{\begin{equation}}
\newcommand{\ee}{\end{equation}}
\newcommand{\bea}{\begin{eqnarray}}
\newcommand{\eea}{\end{eqnarray}}

\renewcommand{\bar}{\overline}

\begin{document} 

\count\footins=600

\title{$\mathcal{PT}$-Symmetric Non-Hermitian Quantum Field Theories\\ with Supersymmetry}

\author{Jean Alexandre}
\email{jean.alexandre@kcl.ac.uk}
\affiliation{Department of Physics, King's College London,\\ 
London WC2R 2LS, United Kingdom}

\author{John Ellis}
\email{john.ellis@cern.ch}
\affiliation{Department of Physics, King's College London,\\ 
London WC2R 2LS, United Kingdom}
\affiliation{National Institute of Chemical Physics \& Biophysics, R\"avala 10, 10143 Tallinn, Estonia}
\affiliation{Theoretical Physics Department, CERN, CH-1211 Geneva 23, Switzerland}

\author{Peter Millington}
\email{p.millington@nottingham.ac.uk}
\affiliation{School of Physics and Astronomy, University of Nottingham,\\ Nottingham NG7 2RD, United Kingdom\vspace{1cm}}

\begin{abstract}

We formulate supersymmetric non-Hermitian quantum field theories with $\mathcal{PT}$ symmetry,
starting with free chiral boson/fermion models and then including trilinear superpotential interactions.
We consider models with both Dirac and Majorana fermions, analyzing them in terms of superfields and
at the component level. We also discuss the relation between the equations of motion, the (non-)invariance
of the Lagrangian and the (non-)conservation of the supercurrents in the two models.
We exhibit a similarity transformation
that maps the free-field supersymmetric $\mathcal{PT}$-symmetric Dirac model to a supersymmetric
Hermitian theory, but there is, in general, no corresponding similarity transformation for the Majorana model.
In this model, we find generically a mass splitting between bosons and fermions,
even though its construction is explicitly supersymmetric, offering a novel non-Hermitian mechanism for soft supersymmetry breaking.\\
~~\\
~~\\
KCL-PH-TH/2020-01, CERN-TH-2020-008
~~\\
April 2020
\\
\noindent\footnotesize{This is an author-prepared post-print of \href{https://doi.org/10.1103/PhysRevD.101.085015}{Phys.\ Rev.\ D {\bf 101} (2020) 085015}, published by the American Physical Society under the terms of the \href{https://creativecommons.org/licenses/by/4.0/}{CC BY 4.0} license (funded by SCOAP\textsuperscript{3}).}
\end{abstract}

\maketitle


\section{Introduction}

Conventional quantum mechanics and quantum field theory are formulated using Hermitian Hamiltonians and
Lagrangians, respectively. However, in recent years there has been increasing interest in extensions
to non-Hermitian quantum theories~\cite{Bender:2002vv}, 
particularly those with $\mathcal{PT}$ symmetry~\cite{Bender:1998ke, Bender:2005tb}, which have real spectra and find
applications in many areas such as optonics~\cite{Longhi, El-Ganainy} and phase transitions~\cite{Ashida, Matsumoto:2019are}. It has also been suggested that non-Hermitian quantum field 
theory might also have applications in fundamental physics, e.g., to 
neutrino physics~\cite{JonesSmith:2009wy, Alexandre:2015kra, Alexandre:2017fpq, Ohlsson:2015xsa}, dark matter~\cite{Rodionov:2017dqt}, 
Higgs decays~\cite{Korchin:2016rsf} and particle mixing~\cite{Pilaftsis:1997dr}. 
It has been shown that it is possible to carry over to $\mathcal{PT}$-symmetric non-Hermitian theories
familiar concepts from Hermitian quantum field theory such as the spontaneous
breaking of global symmetries~\cite{AEMS1, Mannheim:2018dur, Fring:2019hue} and the Englert-Brout-Higgs mechanism in gauge theories~\cite{Mannheim:2018dur, AEMS2, Millington:2019dfn, AEMS3, Fring:2019xgw}, despite
the appearance of subtleties~\cite{alexandre2017symmetries,Alexandre:2017erl} in the relationship
between current conservation, Lagrangian symmetries and Noether's theorem~\cite{Noether} in  non-Hermitian
$\mathcal{PT}$-symmetric theories.

Supersymmetry~\cite{WZ} is a very attractive framework within the conventional Hermitian quantum field theory paradigm,
as it plays a key role in string theory and may play interesting phenomenological roles by stabilising 
the hierarchies of mass scales~\cite{hierarchy}, providing a candidate for dark matter~\cite{LSP}, aiding the grand unification of
gauge couplings~\cite{GUTs} and stabilising the electroweak vacuum~\cite{EDRoss}. Moreover, approximate supersymmetry emerges in a number of less fundamental
physical systems in optonics~\cite{optonics}, condensed-matter physics~\cite{CMSUSY}, 
atomic physics and nuclear physics~\cite{Iachello}. Hence, it is interesting
to explore whether and how the framework of supersymmetry can be extended to $\mathcal{PT}$-symmetric 
non-Hermitian quantum field theories, as we do here for the first time in $3+1$ dimensions.~\footnote{See Ref.~\cite{Bender:1997ps} for a pioneering discussion in $1+1$ dimensions. We note also that the appearance of supersymmetry in a $\mathcal{PT}$-symmetric quantum-mechanical model was discovered in Refs.~\cite{Znojil:2000fr, DDT}.
Relations between Hermitian theories and
$\mathcal{PT}$-symmetric non-Hermitian theories have been derived in the framework of supersymmetric quantum
mechanics~\cite{SQM}, see Refs.~\cite{ZCR, Koohrokhi} and references therein.}

We start by considering $\mathcal{PT}$-symmetric non-Hermitian theories with free bosons and fermions,
studying whether they accommodate supersymmetry, as is the case for free Hermitian theories. We recall that  a necessary condition for supersymmetry is that
the fermionic and bosonic mass spectra coincide. This is not a trivial issue, since a non-Hermitian
fermionic mass term $\propto \bar{\psi} \gamma^5\psi$ is possible for a single species of fermion, whereas a non-Hermitian
bosonic squared-mass term is possible only if there are at least two complex bosons: $\propto \phi_a^\star \phi_b - \phi_b^\star \phi_a$. We discuss the construction of $\mathcal{PT}$-symmetric supersymmetric theories  with a pair of chiral superfields, using the
superfield representation and an appropriate superpotential,  examining
the conditions for the mass spectra to be real and identical, and discussing the extension to
interacting theories. We present our discussion in two formulations of the fermionic sector, 
one in terms of Dirac fermions and the other in terms of Majorana fermions. 

We discuss the model with Dirac fermions in Section~\ref{sec:Dirac}, constructing the $\mathcal{PT}$-symmetric non-Hermitian free-particle model
with Dirac fermions in Section~\ref{sec:FreeDirac} and showing in Section~\ref{sec:Similarity} how it can be related by a similarity transformation~\cite{BJR} to a free-particle Hermitian supersymmetric model. We then discuss supersymmetry transformations in Section~\ref{sec:Supercurrents},
introducing four possible definitions of the supercurrent and discussing the corresponding (non-)invariance properties of the Lagrangian and the (non-)conservation of the corresponding supercurrents. Non-Hermitian dimension-3 bosonic interactions are introduced in Section~\ref{sec:Interactions}. The free-particle model with Majorana fermions is discussed in Section~\ref{sec:Majorana}, initially in its component
representation in Section~\ref{sec:Components} and then in its superfield representation in Section~\ref{sec:MajoranaSuperfields}, after which we discuss the particle spectrum of the Majorana model in Section~\ref{sec:MajoranaSpectrum}. We look for a similarity transformation to a Hermitian model in Section~\ref{sec:MajoranaSimilarity}, finding that it is not possible, in general, to map the
non-Hermitian Majorana model to a supersymmetric Hermitian one.
Supersymmetry transformations and the supercurrent are discussed in Section~\ref{sec:MajoranaSupercurrent}.
Finally, in Section~\ref{sec:conx}, we discuss our conclusions and mention some directions for future work.

\section{$\mathcal{PT}$-symmetric non-Hermitian Supersymmetric Model with Dirac Fermions}
\label{sec:Dirac}

\subsection{Free-Particle Model Construction}
\label{sec:FreeDirac}

The minimal model we consider contains two $\mathcal{N}=1$ scalar chiral superfields $\Phi_a: a=1,2$,
which can be written as follows in conventional notation:
\bea
\Phi_a&=&\phi_a+\sqrt2\theta\chi_a+\theta\theta F_a-i\theta\sigma^\nu\theta^{\dag}\partial_\nu\phi_a\nonumber\\
&&+\frac{i}{\sqrt2}\theta\theta\partial_\nu\chi_a\sigma^\nu\theta^{\dag}-\frac{1}{4}\theta\theta\theta^{\dag}\theta^{\dag}\Box\phi_a~,
\label{Msuper}
\eea
where the $\phi_{1,2}$ are the complex scalar components of the superfields, the $\chi_{1,2}$ are two-component Weyl fermions, the $F_{1,2}$ are complex auxiliary fields, and $\theta^{\alpha}$ and $\theta^{\dag}_{\dot{\alpha}}$ are Grassmann variables. Assuming a minimal K\"ahler potential, the kinetic part $\mathcal{L}_{K}$ of the corresponding free Lagrangian can be written in the usual way as
\bea\label{Lkin}
\mathcal{L}_{K}&=&\int {\rm d}^2\theta^\dagger {\rm d}^2\theta \left(|\Phi_1|^2+|\Phi_2|^2\right)\nonumber\\
&=&\partial_{\nu}\phi^{\dagger}_a\partial^{\nu}\phi_a+i\chi_{a,\dot{\alpha}}^{\dag}\bar{\sigma}^{\nu\dot{\alpha}\beta}\partial_{\nu}\chi_{a,\beta}+F_a^{\dagger} F_a~,
\eea
up to surface terms. One can construct a free-field $\mathcal{PT}$-symmetric model by postulating
the following non-Hermitian combination of superpotential terms:
\begin{equation}
\mathcal{L}_{W,{\rm Dirac}}=\int{\rm d}^2\theta\;W(1-\xi)+\int{\rm d}^2\theta^{\dag}\;W^{\dagger}(1+\xi) \, ,
\label{nonHermW}
\end{equation}
where $\xi$ is a real parameter, with
\begin{equation}
W=m\Phi_1\Phi_2~,
\end{equation}
which yields
\begin{equation}
\mathcal{L}_{W,{\rm Dirac}}=m(1-\xi)(\phi_1F_2+F_1\phi_2-\chi_1^{\alpha}\chi_{2,\alpha})+m(1+\xi)(F_2^{\dagger}\phi_1^{\dagger}+\phi_2^{\dagger} F_1^{\dagger}-\chi^{\dag}_{2,\dot{\alpha}}\chi^{\dag\dot{\alpha}}_{1})~.
\label{LW}
\end{equation}

Due to the non-Hermiticity of the Lagrangian,
\begin{equation}
\mathcal{L}_{\rm Dirac}=\mathcal{L}_K+\mathcal{L}_{W,{\rm Dirac}},
\end{equation}
we have that
\begin{equation}
\label{eq:ELs}
\frac{\partial\mathcal{L}_{\rm Dirac}}{\partial F^{\dagger}_a}=F_a+m(1+\xi)\phi_{\slashed{a}}^{\dagger}=0 \qquad \slashed{\Leftrightarrow} \qquad \frac{\partial\mathcal{L}_{\rm Dirac}}{\partial F_a}=F_a^{\dagger}+m(1-\xi)\phi_{\slashed{a}}=0~,
\end{equation}
except for trivial solutions, where we use the notation $\slashed{1} \equiv 2$ and $\slashed{2} \equiv 1$. It would therefore appear that there is a four-fold ambiguity in the choice of on-shell condition for the auxiliary fields $F_1$ and $F_2$.  However, as first identified in Ref.~\cite{alexandre2017symmetries}, we are, in fact, free to choose any one of the Euler-Lagrange equations to define the equations of motion; each choice leads to the same physics.  In the present case, we can readily convince ourselves that any choice leads to the same Lagrangian for the remaining scalar and fermionic fields:
\begin{align}
\label{eq:LDiracOS}
\mathcal{L}_{\rm Dirac}^{\rm OS}&=\partial_{\nu}\phi_a^{\dagger}\partial^{\nu}\phi_a-m^2(1-\xi^2)|\phi_a|^2\nonumber\\&\qquad+i\chi^{\dag}_{a,\dot{\alpha}}\bar{\sigma}^{\nu\dot{\alpha}\beta}\partial_{\nu}\chi_{a,\beta}-m(1-\xi)\chi_1^{\alpha}\chi_{2,\alpha}-m(1+\xi)\chi^{\dag}_{2,\dot{\alpha}}\chi^{\dag\dot{\alpha}}_1~,
\end{align}
where the superscript ``OS'' indicates that the auxiliary fields have been evaluated on-shell. Alternatively, we could have arrived at Eq.~\eqref{eq:LDiracOS} directly and unambiguously via the path integral by functionally integrating over the auxiliary field, as shown in the Appendix.

Choosing the equations of motion for the scalar and fermion fields by varying with respect to $\phi_a^{\dag}$ and $\chi_a^{\dag}$, respectively, we have
\begin{subequations}
\label{eq:DModEOM}
\begin{gather}
\Box\phi_a+m^2(1-\xi^2)\phi_a=0~,\\
i\bar{\sigma}^{\nu\dot{\alpha}\beta}\partial_{\nu}\chi_{a,\beta}-m(1+\xi)\chi_{\slashed{a}}^{\dag\dot{\alpha}}=0~,
\end{gather}
\end{subequations}
along with their Hermitian conjugates.

The pair of two-component Weyl fermions can be combined into a canonically normalised four-component Dirac fermion
\begin{equation}
\psi=\begin{pmatrix} \chi_{2,\alpha} \\ \chi^{\dag\dot{\alpha}}_{1}\end{pmatrix}~,
\end{equation}
in terms of which the Lagrangian takes the form
\begin{equation}
\mathcal{L}_{\rm Dirac}^{\rm OS}=\partial_{\nu}\phi_a^{\dagger}\partial^{\nu}\phi_a-(m^2-\mu^2)|\phi_a|^2+\bar{\psi}i\slashed{\partial}\psi-m\bar{\psi}\psi-\mu\bar{\psi}\gamma_5\psi~,
\end{equation}
where we have defined $\mu \equiv m\xi$ and the gamma matrices are understood in the Weyl basis:
\begin{equation}
\gamma^0=\begin{pmatrix} 0_2 & I_2 \\ I_2 & 0_2 \end{pmatrix}~,\qquad \gamma^i=\begin{pmatrix} 0_2 & \sigma^i \\ -\sigma^i & 0_2 \end{pmatrix}~,\qquad \gamma^5=\begin{pmatrix} -I_2 & 0_2 \\ 0_2 & I_2 \end{pmatrix}~,
\end{equation}
in which $\sigma^i$ ($i=1,2,3$) are the Pauli matrices. The four scalar and four fermion degrees of freedom all have the same squared mass eigenvalues
\begin{equation}
M^2=m^2-\mu^2~,
\end{equation}
manifesting supersymmetry at the level of the mass spectrum.

Considering only the transformations of the $c$-number fields, \footnote{More generally, the viability of non-Hermitian theories may rely on the existence of a discrete anti-linear symmetry of the Hamiltonian~\cite{Mannheim:2015hto}.} the Lagrangian is $\mathcal{PT}$-symmetric if we take~\cite{alexandre2017symmetries}
\begin{subequations}
\label{eq:fermionPT}
\begin{align}
\mathcal{P}:\qquad &\psi(t,\mathbf{x})\to \psi'(t,-\mathbf{x})=P\psi(t,\mathbf{x})~,\nonumber\\
&\bar{\psi}(t,\mathbf{x})\to \bar{\psi}'(t,-\mathbf{x})=\bar{\psi}(t,\mathbf{x})P~,\\
\mathcal{T}:\qquad &\psi(t,\mathbf{x})\to \psi'(-t,\mathbf{x})=T\psi^*(t,\mathbf{x})~,\nonumber\\
&\bar{\psi}(t,\mathbf{x})\to \bar{\psi}'(-t,\mathbf{x})=\bar{\psi}^*(t,\mathbf{x})T~,
\end{align}
\end{subequations}
where $P=\gamma^0$ and $T=i\gamma^1\gamma^3$ in $3+1$ dimensions. The anti-Hermitian mass term is then odd under both $\mathcal{P}$ and $\mathcal{T}$. We note that the eigenvalues are independent of the sign of $\mu$, and that the eigenvalues are real when $|\mu| < |m|$, in which case the model
is in the unbroken phase of $\mathcal{PT}$ symmetry.

Exceptional points occur at $\mu=\pm m$, corresponding to $\xi = \pm 1$. In these cases, the theory becomes massless and we lose either the left- or the right-chiral Weyl fermion on-shell~\cite{Alexandre:2015kra, Alexandre:2017fpq}. We note that, by virtue of the supersymmetry, the scalar sector inherits the masslessness in spite of having an entirely Hermitian Lagrangian. In addition, beyond the exceptional point in the $\mathcal{PT}$-broken phase, where $|\xi| > 1$ and $|\mu| > |m|$, both the scalar and fermion mass eigenspectra become complex. The scalar sector inherits the $\mathcal{PT}$ phase transition from the fermion sector by virtue of the supersymmetry. 

\subsection{Mapping to a Hermitian Theory via a Similarity Transformation}
\label{sec:Similarity}

The Lagrangian in Eq.~\eqref{eq:LDiracOS} can be mapped to that of a Hermitian theory by the following similarity transformation~\cite{BJR}:
\begin{equation}
\mathcal{L}^{\rm OS}_{\rm Dirac} \to \mathcal{L}_{\rm Dirac}^{{\rm OS}\prime}=S\mathcal{L}_{\rm Dirac}^{\rm OS}S^{-1}~,\qquad \mathcal{L}_{\rm Dirac}^{\rm OS\prime}=\left(\mathcal{L}_{\rm Dirac}^{{\rm OS}\prime}\right)^{\dag}~,
\end{equation}
with
\begin{equation}
S=\exp\left[-{\rm arctanh}\,\xi\int{\rm d}^3\mathbf{x}\left(\chi_1^{\dag}(t,\mathbf{x})\chi_1(t,\mathbf{x})+\chi_2^{\dag}(t,\mathbf{x})\chi_2(t,\mathbf{x})\right)\right]~.
\end{equation}
Noting that (wherein there is no summation over $a$ and $b$)
\begin{subequations}
\begin{align}
\int{\rm d}^3\mathbf{y}\Big[\chi_a^{\dag}(t,\mathbf{y})\chi_a(t,\mathbf{y}),\chi_a(t,\mathbf{x})\chi_b(t,\mathbf{x})\Big]&=-(1+\delta_{ab})\chi_a(t,\mathbf{x})\chi_b(t,\mathbf{x})~,\\
\int{\rm d}^3\mathbf{y}\Big[\chi_a^{\dag}(t,\mathbf{y})\chi_a(t,\mathbf{y}),\chi_b^{\dag}(t,\mathbf{x})\chi_a^{\dag}(t,\mathbf{x})\Big]&=(1+\delta_{ab})\chi_b^{\dag}(t,\mathbf{x})\chi_a^{\dag}(t,\mathbf{x})~,\\
\int{\rm d}^3\mathbf{y}\Big[\chi_a^{\dag}(t,\mathbf{y})\chi_a(t,\mathbf{y}),\chi_b^{\dag}(t,\mathbf{x})\bar{\sigma}\cdot\partial\chi_b(t,\mathbf{x})\Big]&=0~,
\end{align}
\end{subequations}
and using the identities
\begin{subequations}
\begin{align}
\sum_{n\,=\,0}^{\infty}\frac{1}{n!}\left(\pm {\rm arctanh}\,\xi\right)^n=\exp\left(\pm {\rm arctanh}\,\xi\right)=\left(\frac{1\pm\xi}{1\mp\xi}\right)^{1/2}~,
\end{align}
\end{subequations}
we then find
\begin{align}
\label{eq:DiracLprime}
\mathcal{L}^{{\rm OS}\prime}_{\rm Dirac}&=\partial_{\nu}\phi_a^{\dagger}\partial^{\nu}\phi_a-m^2(1-\xi^2)|\phi_a|^2\nonumber\\&\qquad+i\chi^{\dag}_{a,\dot{\alpha}}\bar{\sigma}^{\nu\dot{\alpha}\beta}\partial_{\nu}\chi_{a,\beta}-m\sqrt{1-\xi^2}\left(\chi_1^{\alpha}\chi_{2,\alpha}+\chi^{\dag}_{2,\dot{\alpha}}\chi^{\dag\dot{\alpha}}_1\right)~,
\end{align}
which is Hermitian, as required.  We note that this Lagrangian is isospectral to the original non-Hermitian one.

The Lagrangian in Eq.~\eqref{eq:DiracLprime} can be expressed (off-shell) in terms of chiral superfields as
\begin{equation}
\mathcal{L}'_{\rm Dirac}=\mathcal{L}_{K}+\sqrt{1-\xi^2}\left[\int{\rm d}^2\theta\,W+\int{\rm d}^2\theta^{\dag}\,W^{\dag}\right]~,
\end{equation}
and we obtain
\begin{align}
\mathcal{L}'_{\rm Dirac}&=\partial_{\nu}\phi_a^{\dagger}\partial^{\nu}\phi_a+i\chi^{\dag}_{a,\dot{\alpha}}\bar{\sigma}^{\nu\dot{\alpha}\beta}\partial_{\nu}\chi_{a,\beta}+F_a^{\dagger}F_a\nonumber\\&+m\sqrt{1-\xi^2}\left(\phi_aF_{\slashed{a}}-\chi_1^{\alpha}\chi_{2,\alpha}+F_{\slashed{a}}^{\dagger}\phi_a^{\dagger}-\chi^{\dag}_{2,\dot{\alpha}}\chi^{\dag\dot{\alpha}}_{1}\right)~.
\end{align}
For the Hermitian Lagrangian, there is no ambiguity in choosing the on-shell condition for the auxiliary fields, which are
\begin{equation}
F_a=-m\sqrt{1-\xi^2}\phi_{\slashed{a}}^{\dagger}~,
\end{equation}
and we immediately recover the Lagrangian in Eq.~\eqref{eq:DiracLprime}.

\if
The pair of two-component Weyl fermions can be combined into a canonically-normalised four-component Dirac fermion
\begin{equation}
\psi=\sqrt{2}\begin{pmatrix} \chi^{\alpha}_2 \\ \chi^{\dag}_{1,\dot{\alpha}}\end{pmatrix}~,
\end{equation}
in terms of which the Lagrangian takes the form
\begin{equation}
\mathcal{L}=\partial_{\alpha}\phi_a^\star\partial^{\alpha}\phi_a-(m^2-\mu^2)|\phi_a|^2+\bar{\psi}i\slashed{\partial}\psi-m\bar{\psi}\psi-\mu\bar{\psi}\gamma_5\psi~,
\end{equation}
where we have defined $\mu \equiv m\xi$. 
The four scalar and four fermion degrees of freedom all have the same squared mass eigenvalues
\begin{equation}
M^2=m^2-\mu^2~,
\end{equation}
manifesting supersymmetry at the level of the mass spectrum. We note that the eigenvalues are independent
of the sign of $\mu$, and that the eigenvalues are real when $|\mu| < m$, in which case the model
is $\mathcal{PT}$-symmetric. In the following we work with $\mu > 0$.

An exceptional point occurs at $\mu=\pm m$, corresponding to $\xi = 1$. In this case, the theory becomes massless and we lose either the left- or the right-chiral Weyl fermion on-shell. Notice that, by virtue of the supersymmetry, the scalar sector inherits the masslessness in spite of having an entirely Hermitian Lagrangian. Notice in addition that, beyond the exceptional point in the $\mathcal{PT}$-broken phase, where $\xi > 1$ and $\mu > m$, both the scalar and fermion mass eigenspectra become complex. The scalar sector inherits the $\mathcal{PT}$ phase transition from the fermion sector. 
\fi

\subsection{Supersymmetry Transformations and Supercurrents}
\label{sec:Supercurrents}

Turning to the supersymmetry transformations, we can readily confirm that the Lagrangian given by Eqs.~\eqref{Lkin} and~\eqref{LW} is invariant under the following transformations, up to total derivatives:
\begin{subequations}
\label{eq:SUSYtransfo}
\begin{gather}
\delta \phi_a=\sqrt{2}\epsilon^{\alpha} \chi_{a,\alpha}~,\qquad ~\delta \phi_a^{\dagger}=\sqrt{2}\epsilon^{\dag}_{\dot{\alpha}}\chi^{\dag\dot{\alpha}}_a~,\\
\delta \chi_{a,\alpha}=\sqrt{2}\epsilon_{\alpha} F_a-\sqrt{2}i(\sigma^{\nu}\epsilon^{\dag})_{\alpha}\partial_{\nu}\phi_a~,\qquad \delta \chi^{\dag}_{a,\dot{\alpha}}=\sqrt{2}\epsilon^{\dag}_{\dot{\alpha}}F^{\dagger}_a+\sqrt{2}i(\epsilon\sigma^{\nu})_{\dot{\alpha}}\partial_{\nu}\phi_a^{\dagger}~,\\
\delta F_a=-\sqrt{2}i(\sigma^{\nu} \epsilon^{\dagger})^{\alpha}\partial_{\nu}\chi_{a,\alpha}~,\qquad \delta F^{\dagger}_a=-\sqrt{2}i(\epsilon \sigma^{\nu} )_{\dot{\alpha}}\partial_{\nu}\chi_{a}^{\dagger\dot{\alpha}}~.
\end{gather}
\end{subequations}
Specifically, we obtain
\begin{align}
\delta \mathcal{L}_{\rm Dirac}&=-i\sqrt{2}\partial_{\nu}\left\{\epsilon^{\alpha}\sigma^{\nu}_{\alpha\dot{\beta}}\chi^{\dag\dot{\beta}}_{a}\left[F_a+m(1+\xi)\phi_{\slashed{a}}^{\dag}\right]\right.\nonumber\\&\qquad\left.
+\epsilon^{\dag}_{\dot{\alpha}}\left[-i\chi^{\dagger\dot{\alpha}}_{a}\partial^{\nu}\phi_a+m(1-\xi)\bar{\sigma}^{\nu\dot{\alpha}\beta}\chi_{\slashed{a},\beta}\phi_a\right]\right\}~.
\end{align}
This is as we would expect, given that Eqs.~\eqref{Lkin} and~\eqref{LW} are constructed, respectively, from D and F terms. The corresponding supercurrent is
\begin{align}
J^{\nu}_{\rm Dirac}&=\sqrt{2}\epsilon^{\alpha}\left[\sigma^{\rho}_{\alpha\dot{\beta}}\bar{\sigma}^{\nu\dot{\beta}\gamma}\chi_{a,\gamma}\partial_{\rho}\phi_a^{\dag}+im(1+\xi)\sigma^{\nu}_{\alpha\dot{\beta}}\chi^{\dag\dot{\beta}}_{\slashed{a}}\phi_a^{\dag}\right]\nonumber\\&\qquad+\sqrt{2}\epsilon^{\dag}_{\dot{\alpha}}\left[\bar{\sigma}^{\rho\dot{\alpha}\beta}\sigma^{\nu}_{\beta\dot{\gamma}}\chi^{\dag\dot{\gamma}}_{a}\partial_{\rho}\phi_a+im(1-\xi)\bar{\sigma}^{\nu\dot{\alpha}\beta}\chi_{\slashed{a},\beta}\phi_a\right]~.
\end{align}
This current is not Hermitian, and is not conserved{, except in the Hermitian limit $\xi\to 0$. Specifically, using the equations of motion in Eq.~\eqref{eq:DModEOM}, we find}
\begin{equation}
{\partial_{\nu}J^{\nu}_{\rm Dirac}=\sqrt{2}\epsilon^{\alpha}\left[2m^2\xi(1+\xi)\chi_{a,\alpha}\phi_a^{\dag}\right]+\sqrt{2}\epsilon^{\dag}_{\dot{\alpha}}\left[-2im\xi\bar{\sigma}^{\nu\dot{\alpha}\beta}\chi_{\slashed{a},\beta}\partial_{\nu}\phi_a\right] \neq  0~.}
\end{equation}
The latter is, however, not unexpected, since we know that conserved currents are not related to transformations that leave the Lagrangian invariant in the case of non-Hermitian theories, see Ref.~\cite{alexandre2017symmetries}.~\footnote{We remark that we are working here with the field variables and not their expectation values; since, as illustrated in the Appendix, we have that $\braket{\mathcal{O}}\neq \braket{\mathcal{O}^{\dag}}^*$ in general for non-Hermitian theories, the divergence of the expectation value of the current may still vanish. We leave further study of this, and the subtleties of the classical limit/background-field method for non-Hermitian quantum field theories, to future work.}

We have seen already that there is a four-fold freedom in choosing the on-shell condition for the auxiliary fields $F_a$. While each choice leads to the same Lagrangian, it is clear from Eq.~\eqref{eq:SUSYtransfo} that these choices lead to distinct supersymmetry transformations. In general, and as we will show, there are 16 possible sets of supersymmetry transformations, which we summarize as follows by introducing the independent parameters $s_a,\bar{s}_a=\pm 1$ for $a=1,2$:
\begin{subequations}
\begin{gather}
\delta \phi_a=\sqrt{2}\epsilon^{\alpha} \chi_{a,\alpha}~,\qquad ~\delta \phi_a^{\dagger}=\sqrt{2}\epsilon^{\dag}_{\dot{\alpha}}\chi^{\dag\dot{\alpha}}_a~,\\
\delta \chi_{a,\alpha}=-\sqrt{2}\left[\epsilon_{\alpha}m(1+s_a\xi)\phi_{\slashed{a}}^{\dagger}+i(\sigma^{\nu}\epsilon^{\dag})_{\alpha}\partial_{\nu}\phi_a\right]~,\nonumber\\ \delta \chi^{\dag}_{a,\dot{\alpha}}=-\sqrt{2}\left[\epsilon^{\dag}_{\dot{\alpha}}m(1+\bar{s}_a\xi)\phi_{\slashed{a}}-i(\epsilon\sigma^{\nu})_{\dot{\alpha}}\partial_{\nu}\phi_a^{\dagger}\right]~.
\end{gather}
\end{subequations}

The variation of the Lagrangian under these transformations is
\begin{align}
\delta \mathcal{L}^{\rm OS}_{\rm Dirac}&=\sqrt{2}\epsilon^{\alpha}\left\{-im\xi(1-s_a)\sigma^{\nu}_{\alpha\dot{\beta}}\chi_a^{\dag\dot{\beta}}\partial_{\nu}\phi_{\slashed{a}}^{\dag}-m^2\xi(1-\xi)(1-s_{\slashed{a}})\chi_{a,\alpha}\phi_a^{\dag}\right\}\nonumber\\&+\sqrt{2}\epsilon^{\dag}_{\dot{\alpha}}\left\{\partial_{\nu}\left[\chi_a^{\dag\dot{\alpha}}\partial^{\nu}\phi_a-im\bar{\sigma}^{\nu\dot{\alpha}\beta}\chi_{a,\beta}\phi_{\slashed{a}}\right]\right.\nonumber\\&\qquad\left.+im\xi\bar{\sigma}^{\nu\dot{\alpha}\beta}\left[\chi_{a,\beta}(\partial_{\nu}\phi_{\slashed{a}})-\bar{s}_a(\partial_{\nu}\chi_{a,\beta})\phi_{\slashed{a}}\right]+m^2\xi(1+\xi)(1+\bar{s}_{\slashed{a}})\chi_a^{\dag\dot{\alpha}}\phi_a\right\}~.
\end{align}
We see that this reduces to a total derivative: (i) in the Hermitian limit $\xi\to 0$ and (ii) for $s_a=+1$ and $\bar{s}_a=-1$. The latter case corresponds to making the following replacements in the off-shell transformations in Eq.~\eqref{eq:SUSYtransfo}:
\begin{subequations}
\begin{align}
F_a&\to \braket{F_a}=-m(1+\xi)\phi_{\slashed{a}}^{\dag}~,\\
F_a^{\dag}&\to \braket{F_a^{\dag}}=-m(1-\xi)\phi_{\slashed{a}}~,
\end{align}
\end{subequations}
where we reiterate that $\braket{F_a}\neq \braket{F_a^{\dag}}^*$, see the Appendix.

\subsection{Extension to Include Interactions}
\label{sec:Interactions}

It is possible to extend the non-Hermitian superpotential (\ref{nonHermW}) to include interactions by adding trilinear terms:
\begin{equation}
\Delta \mathcal{L}_{W,{\rm Dirac}} = \int{\rm d}^2\theta\; W_I+\int{\rm d}^2\theta^{\dag}\;W^{\dag}_I ~,
\label{WI}
\end{equation}
where $W_I$ is an arbitrary third-order polynomial function of $\Phi_{1,2}$, and we have assumed for simplicity
that $\Delta \mathcal{L}_{W,{\rm Dirac}}$ is Hermitian, which is not necessarily the case in general. With this modification,
Eq.~\eqref{LW} acquires extra terms:
\begin{equation}
\Delta \mathcal{L}_{W,{\rm Dirac}} = \frac{ \partial w_I}{\partial \phi_a}F_a
- \frac{\partial^2 w_I}{\partial \phi_a \partial \phi_b} \chi_a^{\alpha}\chi_{b,\alpha}+{\rm H.c.}~,
\label{LWI}
\end{equation}
{where $w_I$ is the same arbitrary third-order polynomial function of the complex scalar fields $\phi_{1,2}$, and summations over the indices $a, b$ are to be understood.} The {two equivalent} extremum conditions {for the auxiliary fields} become
\begin{equation}
 \frac{\partial\mathcal{L}}{\partial F^{\dag}_a}=F_a+m(1 {\pm}\xi)\phi_{\slashed{a}}^{\dag} + \frac{ \partial w^{\dag}_I}{\partial \phi^{\dag}_a} =0~,
\label{newEoM}
\end{equation}
leading to the following {on-shell} Lagrangian:
\begin{align}
\mathcal{L}_{\rm Dirac}^{\rm OS}&=\partial_{\nu}\phi_a^{\dag}\partial^{\nu}\phi_a-m^2(1-\xi^2)|\phi_a|^2\nonumber\\&{+i\chi_{a,\dot{\alpha}}^{\dag}\bar{\sigma}^{\nu\dot{\alpha}\beta}\partial_{\nu}\chi_{a,\beta}-m(1-\xi)\chi_1^{\alpha}\chi_{2,\alpha}-m(1+\xi)\chi_{2,\dot{\alpha}}^{\dag}\chi_{1}^{\dag\dot{\alpha}}}\nonumber\\& {- m(1-\xi)\frac{ \partial w^{\dag}_I}{\partial \phi^{\dag}_{\slashed{a}}} \phi_{a} - m(1+\xi)\phi_{a}^{\dag} \frac{ \partial w_I}{\partial \phi_{\slashed{a}}} - \left|\frac{ \partial w^{\dag}_I}{\partial \phi^{\dag}_a} \frac{ \partial w_I}{\partial \phi_a}\right|}\nonumber\\&{-\frac{\partial^2w_{I}}{\partial\phi_a\partial\phi_b}\chi_a^{\alpha}\chi_{b,\alpha}-\frac{\partial^2w^{\dag}_{I}}{\partial\phi_a^{\dag}\partial\phi_b^{\dag}}\chi_{a,\dot{\alpha}}^{\dag}\chi_{b}^{\dag\dot{\alpha}}}~.
\label{LI}
\end{align}

{We make two key observations: first, the interacting Lagrangian remains independent of the choice of extremum condition for the auxiliary fields, as in the free case; and second, the non-Hermiticity of the free part of the Lagrangian has metastasized into the interactions.~\footnote{{The corollary of this observation is that}, unlike the case of a purely scalar field theory, where non-Hermiticity may be restricted to dimension-2 mass terms, non-Hermiticity in such a supersymmetric field theory cannot be limited to mass terms alone, but must include also dimension-3 terms.} Specifically,} the Lagrangian (\ref{LI}) contains non-Hermitian bosonic  interactions of dimension 3,  whereas the dimension-4 bosonic interactions are Hermitian, as are the dimension-4 fermion-boson Yukawa interactions in Eq.~(\ref{LWI}),
by virtue of the assumption that $\Delta \mathcal{L}_{W,{\rm Dirac}}$ (\ref{WI}) is Hermitian.~We expect that the renormalization properties
of this softly-non-Hermitian model are similar to those of a Hermitian supersymmetric model, i.e., the Lagrangian parameters
undergo wave-function renormalization only. In this case, the non-Hermitian parameters could be naturally small, by analogy
with soft supersymmetry-breaking parameters in a Hermitian supersymmetric model.

Finally, we remark that the similarity transformation in Section~\ref{sec:Similarity} does not map this theory to a Hermitian one. The reasons are two-fold: first, we have that
\begin{subequations}
\begin{align}
S\chi_1^{\alpha}\chi_{2,\alpha}S^{-1}&\to \left(\frac{1+\xi}{1-\xi}\right)^{1/2}\chi_1^{\alpha}\chi_{2,\alpha}~,\\
S\chi_{2,\dot{\alpha}}^{\dag}\chi_{1}^{\dag\dot{\alpha}}S^{-1}&\to \left(\frac{1-\xi}{1+\xi}\right)^{1/2}\chi_{2,\dot{\alpha}}^{\dag}\chi_{1}^{\dag\dot{\alpha}}~,
\end{align}
\end{subequations}
which leaves the Yukawa interactions non-Hermitian; and second, this similarity transformation acts only on the fermion fields and therefore leaves the dimension-3 bosonic interactions non-Hermitian. Any similarity transformation of the interacting theory to a Hermitian one would depend on the specific form of the interactions,~\footnote{For a discussion of similarity transformations linking quantum mechanical systems with different physical properties, see, e.g.,~Refs.~\cite{Inzunza:2019sct, Inzunza:2020ogw}.} and we leave further investigation to future work.

\section{$\mathcal{PT}$-symmetric non-Hermitian Supersymmetric Model with Majorana Fermions}
\label{sec:Majorana}

\subsection{Component Representation}
\label{sec:Components}

We consider first a minimal free-particle model containing two complex scalar fields $\phi_1,\phi_2$ and two Majorana fermions $\psi_1,\psi_2$, 
with mass terms that include both Hermitian and anti-Hermitian mixing {\cite{Alexandre:2015kra, Alexandre:2017fpq, BJR, Beygi:2019qab}}. 
Notice that this amounts to four bosonic and four fermionic degrees of freedom.

The $\mathcal{PT}$-symmetric, non-Hermitian free-boson Lagrangian is ($a=1,2$)
\be\label{scalarmass}
\mathcal{L}_{\rm scal}=\partial_{\nu}\phi^{\dag}_a\partial^{\nu}\phi_a-
\begin{pmatrix} \phi_1^{\dag} & \phi_2^{\dag} \end{pmatrix}
\begin{pmatrix} m_1^2 & \mu_s^2 \\ -\mu_s^{2} & m_2^2 \end{pmatrix}
\begin{pmatrix} \phi_1 \\ \phi_2 \end{pmatrix}~,
\ee
where $m_1^2$, $m_2^2$ and $\mu_s^2$ are real. The eigenvalues of the mass matrix are
\be\label{scalareigenmasses}
M_{s,\pm}^2=\frac{1}{2}(m_1^2+m_2^2)\pm\frac{1}{2}\sqrt{(m_1^2-m_2^2)^2-4\mu_s^4}~,
\ee
and these are real as long as 
\be\label{scalarcondition}
(m_1^2-m_2^2)^2\ge4\mu_s^4~.
\ee
The scalar Lagrangian is $\mathcal{PT}$ symmetric with respect to transformations of the $c$-number fields, if these transform as~\cite{alexandre2017symmetries}
\begin{subequations}
\begin{align}
\mathcal{P}:\qquad &\phi_1(t,\mathbf{x})\to \phi_1'(t,-\mathbf{x})=+\phi_1(t,\mathbf{x})~,\nonumber\\
&\phi_2(t,\mathbf{x})\to \phi_2'(t,-\mathbf{x})=-\phi_2(t,\mathbf{x})~,\\
\mathcal{T}:\qquad &\phi_1(t,\mathbf{x})\to \phi_1'(-t,\mathbf{x})=+\phi_1^*(t,\mathbf{x})~,\nonumber\\
&\phi_2(t,\mathbf{x})\to \phi_2'(-t,\mathbf{x})=+\phi_2^*(t,\mathbf{x})~,
\end{align}
\end{subequations}
i.e., if one of the fields transforms as a scalar and the other as a pseudoscalar.

The spin-zero bilinear combinations of the Majorana fermions have the following Hermiticity properties:
\bea\label{Majoranaproperties}
&&\bar{\psi}_a\psi_b=\bar{\psi}_b\psi_a=(\bar{\psi}_a\psi_b)^\dagger ~~~~~~~~~~~~~~~\longrightarrow~~~~\mbox{Hermitian}~,\\
&&\bar{\psi}_a\gamma^5\psi_b=\bar{\psi}_b\gamma^5\psi_a=-(\bar{\psi}_a\gamma^5\psi_b)^\dagger~~~~\longrightarrow~~~~\mbox{anti-Hemitian}~,\nonumber
\eea
and the $\mathcal{PT}$-symmetric, non-Hermitian free-fermion Lagrangian is 
\begin{align}
\label{fermionmass}
\mathcal{L}_{\rm ferm}&=\frac{1}{2}\bar{\psi}_ai\slashed\partial\psi_a-\frac{1}{2}m_{aa}\bar{\psi}_a\psi_a
-\frac{1}{2}\mu_f\bar{\psi}_1\gamma^5\psi_2-\frac{1}{2}\mu_f\bar{\psi}_2\gamma^5\psi_1~.
\end{align}
The corresponding $c$-number Lagrangian is $\mathcal{PT}$ symmetric with respect to the transformations in Eq.~\eqref{eq:fermionPT}. The fermion mass terms can be written in terms of the conjugate variables $\psi_a,\psi_a^\dagger$ as
\be
-\frac{1}{2}\begin{pmatrix} \psi_1^\dagger & \psi_2^\dagger \end{pmatrix}
\begin{pmatrix} m_{11}\gamma^0 & \mu_f\gamma^0\gamma^5 \\ \mu_f\gamma^0\gamma^5 & m_{22}\gamma^0 \end{pmatrix}
\begin{pmatrix} \psi_1 \\ \psi_2 \end{pmatrix}~,
\ee
and the mass eigenvalues are
\be\label{fermioneigenmasses}
M_{f,\pm}=\frac{1}{2}(m_{11}+m_{22})\pm\frac{1}{2}\sqrt{(m_{11}-m_{22})^2-4\mu_f^2}~,
\ee
up to an overall minus sign. These are real as long as 
\be\label{fermioncondition}
(m_{11}-m_{22})^2\ge4\mu_f^2~.
\ee

\subsection{Superfield Representation}
\label{sec:MajoranaSuperfields}

\if
This minimal model can be written in terms of two $N=1$ scalar chiral superfields ($a=1,2$),
which can be written as follows in conventional notation:
\bea
\Phi_a&=&\phi_a+\sqrt2\theta\chi_a+\theta\theta F_a+i\partial_\nu\phi_a\theta\sigma^\nu\bar{\theta}\\
&&-\frac{i}{\sqrt2}\theta\theta\partial_\nu\chi_a\sigma^\nu\bar{\theta}-\frac{1}{4}\Box\phi_a\theta\theta\bar{\theta}\bar{\theta}~.
\label{Msuper}
\eea
The kinetic part $\mathcal{L}_{\rm kin}$ of the corresponding free Lagrangian can be written in the usual way as
\bea\label{Lkin}
\mathcal{L}_{\rm kin}&=&\int {\rm d}^2\theta^\dagger {\rm d}^2\theta \left(|\Phi_1|^2+|\Phi_2|^2\right)\\
&=&\partial_{\nu}\phi^\star_a\partial^{\nu}\phi_a+2i\chi_{a,\dot{\alpha}}^{\dag}\bar{\sigma}^{\nu\dot{\alpha}\beta}\partial_{\nu}\chi_{a,\beta}+F_a^\star F_a~.\nonumber
\eea
up to surface terms. 
\fi
We use the same two $\mathcal{N}=1$ scalar chiral superfields ($\Phi_a=1,2$) as in the Dirac model.
In order to incorporate mass terms for the fields in the Majorana model, we introduce the following two superpotentials: 
\be\label{MajoranaSuperpotentials}
W_\pm=\frac{1}{2}m_{11}\Phi_1^2\mp \frac{1}{2}(m_{12}+m_{21})\Phi_1\Phi_2+\frac{1}{2}m_{22}\Phi_2^2~,
\ee
where $m_{ab}$ are real and symmetric, and we consider the following non-Hermitian Lagrangian
\be\label{L0mass}
\mathcal{L}_{\rm Maj}=\mathcal{L}_K+\int {\rm d}^2\theta~ W_++\int {\rm d}^2\theta^\dagger ~ W_-^{\dag}~.
\ee
The scalar sector derived from the expressions (\ref{Lkin}) and (\ref{L0mass}) is
\be
\label{eq:Majscal}
\mathcal{L}_{\rm scal}= \partial_{\nu}\phi_a^{\dag}\partial^{\nu}\phi_a+F_a^{\dag} F_a+m_{aa}(\phi_aF_a+\phi_a^{\dag} F_a^{\dag})-m_{a\slashed{a}}\big(\phi_aF_{\slashed{a}}-F^{\dag}_{\slashed{a}}\phi_a^{\dag}\big)~,
\ee
and the fermion sector derived from the same is given by
\be
\label{eq:LMajferm}
\mathcal{L}_{\rm ferm}=i\chi^{\dag}_{a,\dot{\alpha}}\bar{\sigma}^{\nu\dot{\alpha}\beta}\partial_{\nu}\chi_{a,\beta}-\frac{1}{2}m_{aa}(\chi_a^{\alpha}\chi_{a,\alpha}+\chi^{\dag}_{a,\dot{\alpha}}\chi^{\dag\dot{\alpha}}_a)+m_{12}(\chi_1^{\alpha}\chi_{2,\alpha}-\chi_{2,\dot{\alpha}}^{\dag}\chi_1^{\dag\dot{\alpha}})~.
\ee

As in the case of the Dirac model, we have a four-fold freedom in choosing the on-shell conditions for the auxiliary fields, e.g., we might take
\be
\frac{\partial \mathcal{L}_{\rm Maj}}{\partial F_a^{\dag}}=F_a+m_{aa}\phi_a^{\dag}+m_{a\slashed{a}}\phi_{\slashed{a}}^{\dag}=0~.
\ee
However, the result of integrating out the auxiliary fields is unique, and, whichever choice we make, we arrive at the following Lagrangian for the scalar sector:
\bea
\mathcal{L}_{{\rm scal}}&=&\partial_{\nu}\phi_a^{\dag}\partial^{\nu}\phi_a-m_a^2\phi_a^{\dag}\phi_a-\mu_s^2(\phi_1^{\dag}\phi_2-\phi_1\phi_2^{\dag})~,
\eea
where
\begin{subequations}
\label{scalarmasses}
\bea
m_a^2&=&m_{aa}^2-m_{a\slashed{a}}^2~,\\
\mu_s^2&=& m_{12}(m_{22}-m_{11})~.
\eea
\end{subequations}

{Choosing the equations of motion for the scalar and fermion fields by varying with respect to $\phi_a^{\dag}$ and $\chi_a^{\dag}$, respectively, we have
\begin{subequations}
\label{eq:MajEOM}
\begin{gather}
\Box\phi_1+m_1^2\phi_1+\mu_s^2\phi_2=0~,\\
\Box\phi_2+m_2^2\phi_2-\mu_s^2\phi_1=0~,\\
i\bar{\sigma}^{\nu\dot{\alpha}\beta}\partial_{\nu}\chi_{1,\beta}-m_{11}\chi_{1}^{\dag\dot{\alpha}}-m_{12}\chi_2^{\dag\dot{\alpha}}=0~,\\
i\bar{\sigma}^{\nu\dot{\alpha}\beta}\partial_{\nu}\chi_{2,\beta}-m_{22}\chi_{2}^{\dag\dot{\alpha}}-m_{12}\chi_1^{\dag\dot{\alpha}}=0~,
\end{gather}
\end{subequations}
along with their Hermitian conjugates.}

The fermion Lagrangian can be recast in terms of two Majorana fermions
\be
\psi_a\equiv\begin{pmatrix} \chi_{a,\alpha} \\ \chi_{a}^{c\dot{\alpha}} \end{pmatrix}~~~~~(a=1,2)~,
\ee
where the charge-conjugate spinor is $\chi_a^c\equiv -i\sigma^2\chi_a^\star$ and $\sigma^2$ is the second Pauli matrix. Making use of the following dictionary between the Weyl and Majorana fermion bilinears:
\begin{subequations}
\bea
\chi_a^{\alpha}\chi_{b,\alpha}&=&\frac{1}{2}\left(\bar{\psi}_a\psi_b-\bar{\psi}_a\gamma^5\bar{\psi}_b\right) \, ,\\
\chi_{a,\dot{\alpha}}^{\dag}\chi^{\dag\dot{\alpha}}_b&=&\frac{1}{2}\left(\bar{\psi}_a\psi_b+\bar{\psi}_a\gamma^5\psi_b\right)~,
\eea
\end{subequations}
we obtain
\be
\mathcal{L}_{\rm ferm}=\frac{1}{2}\bar{\psi}_ai\slashed{\partial}\psi_a-\frac{1}{2}m_{aa}\bar{\psi}_a\psi_a-\frac{1}{2}m_{12}(\bar{\psi_1}\gamma^5\psi_2+\bar{\psi}_2\gamma^5\psi_1)~,
\ee
 and identifying the latter expression with the fermionic mass terms in the Lagrangian (\ref{fermionmass}), we associate
\be
\mu_f=m_{12}~.
\label{fermionmasses}
\ee
 
\subsection{(Non-)Supersymmetric spectrum} 
\label{sec:MajoranaSpectrum} 
 
Given the expressions (\ref{scalarmasses}) for the scalar mass parameters and (\ref{fermionmasses}) for the fermion mass parameter,
one can check that the eigenvalues (\ref{scalareigenmasses}) and (\ref{fermioneigenmasses}) can be written as
\bea
M_{s,\pm}^2&=&\frac{1}{2}(m_{11}^2+m_{22}^2)-m_{12}^2\nonumber\\
&&\pm\frac{1}{2}\sqrt{(m_{11}^2-m_{22}^2)^2-4m_{12}^2(m_{11}-m_{22})^2}~,
\eea
and
\bea
M_{f,\pm}^2&=&\frac{1}{2}(m_{11}^2+m_{22}^2)-m_{12}^2\nonumber\\
&&\pm\sqrt{(m_{11}^2-m_{22}^2)^2-4m_{12}^2(m_{11}+m_{22})^2}~.\nonumber
\eea
We see immediately that 
\be
M_{s,\pm}^2(m_{11}, -m_{22})=M_{s,\pm}
^2(-m_{11},m_{22})=M_{f,\pm}^2(m_{11},m_{22})~.
\label{comparemasses}
\ee
Hence, although the non-Hermitian Lagrangian itself was written entirely in terms of chiral superfields, the spectrum is \emph{not} supersymmetric, 
except in the limiting cases $m_{11}=0$ or $m_{22}=0$ (or the Hermitian limit $m_{12}=0$).

This non-supersymmetric spectrum was to be expected, in view of the different signs in the
superpotentials (\ref{MajoranaSuperpotentials}). {Indeed, mass terms mix coefficients appearing in the superpotentials $W_+$ and $W_-^\dagger$: 
the equation of motion for the auxiliary fields $F_a$ are obtained from taking functional derivatives with respect to $F_a^\dagger$, 
and therefore involve coefficients from $W_-^\dagger$. The resulting expression for $F_a$ is then inserted in terms arising from $W_+$, hence mixing 
coefficients from $W_+$ and $W_-^\dagger$, and so failing to ensure a supersymmetric spectrum when $W_+\ne W_-$.
However, if we assume that one of the diagonal mass terms vanishes, say $m_{22}=0$, we can understand why a 
supersymmetric spectrum is recovered. For a quadratic superpotential, the mass terms do not depend on the overall sign of the  
superpotential, and physical quantities do not depend on the sign of $m_{12}$.
As a consequence all of the combinations
$(m_{11},m_{12}),~(m_{11},-m_{12}),~(-m_{11},m_{12})$ and $(-m_{11},-m_{12})$ lead to the same spectrum, and Eq.~(\ref{comparemasses}) 
shows that we recover identical scalar and fermionic masses. If one switches on the mass term $m_{22}$ though, 
the above properties are still valid but we are left with an additional  relative physical sign 
between $m_{11}$ and $m_{22}$, and we cannot expect a supersymmetric spectrum anymore.}

In order to give another interpretation of the non-supersymmetric spectrum, we can make the supersymmetry breaking explicit by
implementing a phase rotation of the fermion sector via the unitary transformation
\begin{equation}
\label{eq:phaserotation}
\chi_2\to \tilde{\chi}_2=-i\chi_2~,\qquad \chi_2^{\dag}\to \tilde{\chi}
_2^{\dag}=+i\chi_2^{\dag}~,
\end{equation}
which gives the fermionic Lagrangian
\bea
\tilde{\mathcal{L}}_{\rm ferm}&=&i\chi^{\dag}_{a,\dot{\alpha}}\bar{\sigma}^{\nu\dot{\alpha}\beta}\partial_{\nu}\chi_{a,\beta}-\frac{1}{2}m_{11}(\chi_1^{\alpha}\chi_{1,\alpha}+\chi^{\dag}_{1,\dot{\alpha}}\chi^{\dag\dot{\alpha}}_1)+\frac{1}{2}m_{22}(\chi_2^{\alpha}\chi_{2,\alpha}+\chi^{\dag}_{2,\dot{\alpha}}\chi^{\dag\dot{\alpha}}_2)\nonumber\\&&\qquad -im_{12}(\chi_1^{\alpha}\chi_{2,\alpha}+\chi_{2,\dot{\alpha}}^{\dag}\chi_1^{\dag\dot{\alpha}})~\nonumber\\&=&\frac{1}{2}\bar{\psi}_ai\slashed{\partial}\psi_a-\frac{1}{2}m_{11}\bar{\psi}_1\psi_1+\frac{1}{2}m_{22}\bar{\psi}_2\psi_2-\frac{1}{2}m_{12}(\bar{\psi_1}i\gamma^5\psi_2+\bar{\psi}_2i\gamma^5\psi_1)~.
\eea
Putting back the scalar sector, the spectrum is now supersymmetric, but the Lagrangian itself can no longer be written entirely in terms of chiral superfields.

This supersymmetry breaking is entirely a consequence of the non-Hermiticity. In contrast, the supersymmetry remains unbroken in spite of the non-Hermiticity in the 1+1 dimensional model of Ref.~\cite{Bender:1997ps}.  Had we taken a model with an analogous Hermitian mass mixing, arising from either of the superpotentials,~\footnote{The sign of $m_{12}$ is irrelevant.} i.e., taking
\be
\mathcal{L}_{\rm Maj,Herm}=\mathcal{L}_K+\int {\rm d}^2\theta~ W_{\pm}+\int {\rm d}^2\theta^\dagger ~ W_{\pm}^{\dag}~,
\ee
we would have found that the spectrum was fully supersymmetric, with squared masses given by
\bea
\label{eq:M2Herm}
M^2_{\rm Herm,\pm}&=&\frac{1}{2}(m_{11}^2+m_{22}^2)+m_{12}^2\nonumber\\
&&\pm\frac{1}{2}\sqrt{(m_{11}^2-m_{22}^2)^2+4m_{12}^2(m_{11}+m_{22})^2}~,
\eea
cf.~Eqs.~\eqref{scalareigenmasses} and \eqref{fermioneigenmasses}. We note that the mass spectrum remains sensitive to the relative sign of the diagonal fermion mass terms also for the Hermitian mass mixing, such that the fermion phase rotation in Eq.~\eqref{eq:phaserotation} again leads to a non-supersymmetric model, but where this is manifest in both the Lagrangian and the spectrum.

\subsection{Similarity Transformation}
\label{sec:MajoranaSimilarity}

The scalar part of the Lagrangian can be mapped to that of a Hermitian theory via the following similarity transformation~\cite{Mannheim:2018dur}:
\begin{equation}
\mathcal{L}_{\rm scal} \to \mathcal{L}'_{\rm scal}=S_{\phi}\,\mathcal{L}_{\rm scal}\,S_{\phi}^{-1}~,
\end{equation}
with
\begin{equation}
S_{\phi}=\exp\left[\frac{\pi}{2}\int{\rm d}^3\mathbf{x}\,\left(\pi_2(t,\mathbf{x})\phi_2(t,\mathbf{x})+\pi^{\dag}_2(t,\mathbf{x})\phi_2^{\dag}(t,\mathbf{x})\right)\right]~,
\end{equation}
where $\pi_2(t,\mathbf{x})=\dot{\phi}_2^{\dag}(t,\mathbf{x})$ is the conjugate momentum operator. This transforms
\begin{equation}
\phi_2\to -i\phi_2\qquad \text{and}\qquad \phi_2^{\dag}\to -i\phi_2^{\dag}~,
\end{equation}
leading to
\begin{align}
\mathcal{L}'_{{\rm scal}}&=\partial_{\nu}\phi_1^{\dag}\partial^{\nu}\phi_1-m_1^2|\phi_1|^2-\partial_{\nu}\phi_2^{\dag}\partial^{\nu}\phi_2
+m_2^2|\phi_2|^2+i\mu_s^2(\phi_1^{\dag}\phi_2-\phi_2^{\dag}\phi_1)~.
\end{align}
This Lagrangian can be obtained from 
\begin{align}
\mathcal{L}'_{\rm scal}&= \partial_{\nu}\phi_1^{\dag}\partial^{\nu}\phi_1+F_1^{\dag} F_1-\partial_{\nu}\phi_2^{\dag}\partial^{\nu}\phi_2-F_2^{\dag} F_2\nonumber\\&+m_{11}(\phi_1F_1+ F_1^{\dag}\phi_1^{\dag})-m_{22}(\phi_2F_2+F_2^{\dag}\phi_2^{\dag} )\nonumber\\&+im_{12}\big(\phi_1F_2+\phi_2F_1-F_2^{\dag}\phi_1^{\dag} - F_1^{\dag}\phi_2^{\dag}\big)~,
\end{align}
which itself arises from the supersymmetric Lagrangian with the K\"{a}hler potential
\begin{equation}
\label{eq:LKprime}
\mathcal{L}'_{K}=\int {\rm d}^2\theta^\dagger {\rm d}^2\theta \left(|\Phi_1|^2-|\Phi_2|^2\right)
\end{equation}
and the superpotential
\be
\mathcal{L}_{W,{\rm Maj}}'=\int {\rm d}^2\theta~ W'+\int {\rm d}^2\theta^\dagger ~ W^{\prime\dag
}~,
\ee
where
\begin{equation}
W'=\frac{1}{2}m_{11}\Phi_1^2+\frac{1}{2}i(m_{12}+m_{21})\Phi_1\Phi_2-\frac{1}{2}m_{22}\Phi_2^2~.
\end{equation}
However, the resulting fermionic Lagrangian is
\begin{align}
\label{eq:Lfermprime}
\mathcal{L}'_{\rm ferm}&=i\chi^{\dag}_{1,\dot{\alpha}}\bar{\sigma}^{\nu\dot{\alpha}\beta}\partial_{\nu}\chi_{1,\beta}-i\chi^{\dag}_{2,\dot{\alpha}}\bar{\sigma}^{\nu\dot{\alpha}\beta}\partial_{\nu}\chi_{2,\beta}\nonumber\\&-m_{11}(\chi_1^{\alpha}\chi_{1,\alpha}+\chi^{\dag}_{1,\dot{\alpha}}\chi^{\dag\dot{\alpha}}_1)+m_{22}(\chi_2^{\alpha}\chi_{2,\alpha}+\chi^{\dag}_{2,\dot{\alpha}}\chi^{\dag\dot{\alpha}}_2)-im_{12}(\chi_1^{\alpha}\chi_{2,\alpha}-\chi_{2,\dot{\alpha}}^{\dag}\chi_1^{\dag\dot{\alpha}})~,
\end{align}
which cannot be reached by a similarity transformation of the non-Hermitian Lagrangian in Eq.~\eqref{eq:LMajferm}.

Before concluding this section, we remark on the wrong sign of the kinetic term in Eq.~\eqref{eq:LKprime}. {Whilst  in the context of Hermitian quantization this would lead to negative-norm modes, the presence of $\mathcal{PT}$ symmetry is sufficient to ensure that one can always construct a positive-definite inner product consistent with unitary evolution~\cite{Bender:2002vv}.}

\subsection{Supersymmetry Transformations and Supercurrents}
\label{sec:MajoranaSupercurrent}

We can readily confirm that the Lagrangian composed of Eqs.~\eqref{eq:Majscal} and \eqref{eq:LMajferm} is invariant under the supersymmetry transformations given in Eq.~\eqref{eq:SUSYtransfo} up to total derivatives. Specifically, we find
\begin{align}
\delta \mathcal{L}_{\rm Maj}&=\sqrt{2}\epsilon^{\alpha}\Big\{-i\partial_{\nu}\Big[\sigma^{\nu}_{\alpha\dot{\beta}}\chi_a^{\dag\dot{\beta}}\Big(F_a+m_{aa}\phi_a^{\dag}+m_{a\slashed{a}}\phi_{\slashed{a}}^{\dag}\Big)\Big]\Big\}\nonumber\\&+\sqrt{2}\epsilon^{\dag}_{\dot{\alpha}}\Big\{\partial_{\nu}\Big[\chi_{a}^{\dag\dot{\alpha}}\partial^{\nu}\phi_a-i\bar{\sigma}^{\nu\dot{\alpha}\beta}\chi_{a,\beta}\Big(m_{aa}\phi_a-m_{a\slashed{a}}\phi_{\slashed{a}}\Big)\Big]\Big\}~.
\end{align}
Analogously to the Dirac model, the on-shell Lagrangian is invariant under the transformations in Eq.~\eqref{eq:SUSYtransfo}, again up to total derivatives, as long as we make the replacement
\begin{subequations}
\begin{align}
F_a\to\braket{F_a}=-m_{aa}\phi_a^{\dag}-m_{a\slashed{a}}\phi_{\slashed{a}}^{\dag}~,\\
F_a^{\dag}\to\braket{F_a^{\dag}}=-m_{aa}\phi_a+m_{a\slashed{a}}\phi_{\slashed{a}}~.
\end{align}
\end{subequations}
The corresponding supercurrent is
\begin{align}
J^{\nu}_{\rm Maj}&=\sqrt{2}\epsilon^{\alpha}\Big[\sigma^{\rho}_{\alpha\dot{\beta}}\bar{\sigma}^{\nu\dot{\beta}\gamma}\chi_{a,\gamma}\partial_{\rho}\phi_a^{\dag}+i\sigma^{\nu}_{\alpha\dot{\beta}}\chi_a^{\dag\dot{\beta}}\Big(m_{aa}\phi_a^{\dag}+m_{a\slashed{a}}\phi_{\slashed{a}}^{\dag}\Big)\Big]\nonumber\\&
+\sqrt{2}\epsilon^{\dag}_{\dot{\alpha}}\Big[\bar{\sigma}^{\rho\dot{\alpha}\beta}\sigma^{\nu}_{\beta\dot{\gamma}}\chi_a^{\dag\dot{\gamma}}\partial_{\rho}\phi_a+i\bar{\sigma}^{\nu\dot{\alpha}\beta}\chi_{a,\beta}\Big(m_{aa}\phi_a-m_{a\slashed{a}}\phi_{\slashed{a}}\Big)\Big]~,
\end{align}
which is again neither Hermitian nor conserved. Using the equations of motion in Eq.~\eqref{eq:MajEOM}, along with their Hermitian conjugates, we find that the divergence of the current is
\begin{align}
\partial_{\nu}J^{\nu}_{\rm Maj}=\sqrt{2}\epsilon^{\alpha}\Big[2m_{12}\chi_{a,\alpha}\Big(m_{aa}\phi_{\slashed{a}}^{\dag}+m_{a\slashed{a}}\phi_a^{\dag}\Big)\Big]+\sqrt{2}\epsilon^{\dag}_{\dot{\alpha}}\Big[-2im_{12}\bar{\sigma}^{\nu\dot{\alpha}\beta}\chi_{a,\beta}\partial_{\nu}\phi_{\slashed{a}}\Big]~,
\end{align}
which vanishes, as it should, in the Hermitian limit $m_{12}\to 0$.

\section{Conclusions}
\label{sec:conx}

In this paper, we have constructed $\mathcal{PT}$-symmetric 
$\mathcal{N}=1$ supersymmetric quantum field theories for the first time in 3+1 dimensions. We have presented models incorporating 
a pair of chiral supermultiplets and either Dirac or Majorana fermions. We have shown that the free-field
supersymmetric Dirac model is equivalent via a similarity transformation to a 
Hermitian supersymmetric model, but we have found that there is no such equivalence 
in the general Majorana case. As we have described in both models, there is {an
ambiguity} in the definition of the supercurrent, and we have discussed the (non-)invariance of the Lagrangian and the 
(non-)conservation of the Noether current, which are analogous to the corresponding properties of
$\mathcal{PT}$-symmetric models with purely bosonic symmetries~\cite{alexandre2017symmetries}. 
We have also extended the Dirac
model to include Hermitian trilinear superpotential interactions,~\footnote{The Majorana model
may be extended in a similar way, which we leave for further work.} in which case the model contains
non-Hermitian trilinear bosonic interactions as well as non-Hermitian bilinear terms, whereas the
dimension-4 interactions are Hermitian.

This work is only a first step towards the exploration of $\mathcal{PT}$-symmetric 
supersymmetric quantum field theories. One interesting topic to explore will be the general
structure of $\mathcal{PT}$-symmetric quantum field theories with $\mathcal{N}=1$ supersymmetry,
extending models containing only chiral superfields to models including
vector superfields. Another interesting topic will be the study of possible generalization of models
with rigid $\mathcal{N}=1$ supersymmetry to those with local $\mathcal{N}=1$ supersymmetry,
i.e., supergravity theories. Extensions to $\mathcal{N}>1$ supersymmetric models also warrant attention.

The construction of $\mathcal{PT}$-symmetric supersymmetric quantum field theories is all well and good,
but do they have any practical applications? As was mentioned in the Introduction, non-supersymmetric 
$\mathcal{PT}$-symmetric quantum field theories have found many applications in non-fundamental
areas such as optonics~\cite{Longhi,El-Ganainy}, and approximate supersymmetry also 
emerges in many non-fundamental areas such as optonics, condensed-matter physics, 
atomic and nuclear physics~\cite{optonics,CMSUSY,Iachello}. Might it be possible to find applications
of $\mathcal{PT}$-symmetric supersymmetric quantum field theories in non-fundamental areas such as optonics?
A more ambitious, longer-term hope is that these theories may find applications in fundamental physics. 

\section*{Acknowledgements}

PM would like to thank Silvia Nagy for helpful discussions. The work of JA and JE was supported by the United Kingdom STFC Grant ST/P000258/1, 
and that of JE also by the Estonian Research Council via a Mobilitas Pluss grant. The work of PM was supported by a Leverhulme Trust Research Leadership Award (Grant No.~RL-2016-028).

\appendix

\section{Direct Integration}

Let's take the Dirac model as an example. Assuming constant field configurations, the contribution to the Euclidean path integral from each of the auxiliary fields is (no summation over $a$ implied)
\begin{equation}
\mathcal{I}=\int\mathcal{D}F_a\,\mathcal{D}F_a^*\exp\left[-\mathcal{V}\left(F_a^*F_a+m(1-\xi)\phi_{\slashed{a}}F_{a}+m(1+\xi)\phi_{\slashed{a}}^*F_{a}^*\right)\right]~,
\end{equation}
where $\mathcal{V}$ is the volume of $\mathbb{R}^4$. This is integrable, and we find
\begin{equation}
\label{eq:Iresult}
\mathcal{I}=\mathcal{N}\frac{\pi^2}{\mathcal{V}^2}\exp\left[-\mathcal{V} m^2(1-\xi^2)|\phi_a|^2\right]~,
\end{equation}
where $\mathcal{N}$ is an irrelevant constant that depends on the normalization of the functional measure. We can also obtain Eq.~\eqref{eq:Iresult} by expanding around/setting
\begin{equation}
F_{a}=-m(1+\xi)\phi_{\slashed{a}}^*\qquad \text{or}\qquad F_{a}=-m(1-\xi)\phi_{\slashed{a}}^*~,
\end{equation}
corresponding to setting the variation of the exponent with respect to $F_{\slashed{a}}^*$ to zero or the variation of the exponent with respect to $F_{\slashed{a}}$ to zero, respectively.

In addition, we can calculate the expectation values of $F_a$ and $F_a^{*}$:
\begin{subequations}
\begin{align}
\braket{F_a}&=\frac{1}{\mathcal{I}}\int\mathcal{D}F_a\,\mathcal{D}F_a^*\;F_a\exp\left[-\mathcal{V}\left(F_a^*F_a+m(1-\xi)\phi_{\slashed{a}}F_{a}+m(1+\xi)\phi_{\slashed{a}}^*F_{a}^*\right)\right]\nonumber\\&=-m(1+\xi)\phi_{\slashed{a}}^*~,\\
\braket{F_a^*}&=\frac{1}{\mathcal{I}}\int\mathcal{D}F_a\,\mathcal{D}F_a^*\;F_a^*\exp\left[-\mathcal{V}\left(F_a^*F_a+m(1-\xi)\phi_{\slashed{a}}F_{a}+m(1+\xi)\phi_{\slashed{a}}^*F_{a}^*\right)\right]\nonumber\\&=-m(1-\xi)\phi_{\slashed{a}}~.
\end{align}
\end{subequations}
We see immediately that
\begin{equation}
\braket{F_a}\neq \braket{F_a^*}^*~,
\end{equation}
that is, the vacuum state is not invariant under complex conjugation, i.e.,
\begin{equation}
\braket{F_a} = \braket{\Omega|F_a|\Omega} \neq \braket{\Omega|F_a^*|\Omega}^* = \braket{\Omega^*|F_a|\Omega^*}~.
\end{equation}
In this way, choosing the on-shell condition for $F_a$ is equivalent to choosing whether we work with the vacuum $\Omega$ or $\Omega^*$, that is whether we choose
\begin{equation}
\braket{F_a}\equiv\braket{\Omega|F_a|\Omega}=-m(1+\xi)\phi_{\slashed{a}}^*\text{or} \braket{F_a}\equiv\braket{\Omega^*|F_a|\Omega^*}=-m(1-\xi)\phi_{\slashed{a}}^*~.
\end{equation}
Since these differ only in the sign of the non-Hermitian terms, this choice is irrelevant, as we have seen previously.

\end{document}